\documentclass[aps,prd,amssymb,amsmath,eqsecnum,nofootinbib,onecolumn]{revtex4}
\usepackage{amsmath,amssymb}
\usepackage{mathrsfs}
\usepackage{ulem}
\usepackage{graphicx}

\usepackage{color}

\begin{document}

\title{Hierarchical formulation of the self-gravitating, n-dimensional, charged scalar field in spherical symmetry in affine null formalism}

\author{Laura Bridera$^1$, Emanuel Gallo$^1$,$^2$}
\email{egallo@unc.edu.ar}
\affiliation{$^1$ Universidad Nacional de Córdoba, Facultad de Matemática, Astronomía, Física y Computación, Grupo de Relatividad y Gravitación; Córdoba, Argentina.\\$^2$ Consejo Nacional de Investigaciones Científicas y Técnicas,\\ CONICET, IFEG. Córdoba, Argentina. }

\author{Thomas M\"adler$^{3,4}$}
\email{thomas.maedler@mail.udp.cl}
\affiliation{$^3$ Facultad de Ingeniería, Universidad San Sebastián,
Bellavista 7, Recoleta, Chile} 
\affiliation{$^4$ Facultad de Ingenier\'{i}a y Ciencias, Universidad Diego Portales, Avenida Ej\'{e}rcito
Libertador 441, Casilla 298-V, Santiago, Chile. }
\date{July 2026}

\maketitle

\section*{Abstract}
{We develop an affine-null characteristic formulation of the Einstein-Maxwell system coupled to a charged complex scalar field in $n$-dimensional spherical symmetry. By introducing suitable auxiliary variables, the main field equations are cast into a hierarchical system of radial hypersurface equations, supplemented by a transport equation for the scalar field. We discuss the associated characteristic initial-boundary value problem for asymptotic, vertex and null-boundary configurations, and derive the corresponding asymptotic quantities and balance laws. As a consistency check, we recover the scalar-free Reissner-Nordström-Tangherlini family, including both the non-extremal and extremal branches, directly from the hierarchy. The resulting framework provides a systematic setting for the study of charged scalar dynamics and exact black-hole solutions in higher-dimensional affine-null coordinates.}

\section{Introduction} 

Characteristic methods provide a natural framework for studying the Einstein equations in situations where radiation, horizons and null infinity play a central role \cite{Bondi1962,Sachs1962,Penrose1963,Winicour2012}. Instead of foliating spacetime by spacelike hypersurfaces, one uses null hypersurfaces adapted to the causal propagation of the gravitational and matter fields. This point of view was essential in the original analysis of asymptotic radiation and Bondi mass loss, and it remains one of the basic tools in modern characteristic evolution schemes. In spherical symmetry, coupling gravity to a scalar field provides a simple but non-trivial setting in which genuine dynamical effects are retained. The scalar field supplies the propagating degree of freedom, while the metric variables are determined by hypersurface equations once suitable initial and boundary data are prescribed. This structure is especially transparent in Bondi-Sachs coordinates, where the main equations naturally form a radial hierarchy: given the scalar profile on an initial null hypersurface, the metric functions are obtained by radial integration, and the scalar wave equation updates the data to the next null slice \cite{Bondi1962,Sachs1962,Winicour2012}.

The affine-null formulation offers a complementary approach. Its radial coordinate is an affine parameter along the null generators, rather than the areal radius \cite{Winicour2013}. 
This choice is particularly useful near trapped or marginally trapped regions, because the affine parameter can remain regular where the areal radius ceases to be a good radial coordinate \cite{CrespoOliveiraWinicour2019,Madler:2025ibn,10}. For this reason, affine-null coordinates are well adapted for investigations of black-hole interiors and event horizon crossing evolutions \cite{MadlerBaakeHosseiniWinicour2024,8}. However, the price for this regularity is that the usual Bondi-Sachs hierarchy is obscured: one of the Einstein equations contains {a derivative of the areal distance, $r$ pointing out of the null hypersurface}, and therefore cannot be integrated radially in a direct sequential way using the original variables. A way around this obstruction is to introduce auxiliary variables which absorb the problematic derivatives and restore a hypersurface hierarchy. This strategy has been developed in four-dimensional affine-null formulations in different situations \cite{Winicour2013,madler-gallo-conf,CrespoOliveiraWinicour2019,MadlerBaakeHosseiniWinicour2024}. 
{Alternative Bondi-Sachs modified schemes were developed by Gundlach and collaborators \cite{Gundlach:2024mld}, who applied it to the study of critical phenomena \cite{Gundlach:2024eds} as well as to evolutions penetrating the event horizon \cite{Gundlach:2025mys}.}

The aim of the present work is to extend the construction {of \cite{3}} to $n$-dimensional spherical symmetry. We consider the Einstein-Maxwell system coupled to a charged massless complex scalar field and show that, after introducing suitable variables, the main equations can again be written as a sequential system of radial hypersurface equations, supplemented by a first-order transport equation for the scalar field.

The resulting hierarchy has several useful features. 
First, it provides a characteristic initial-boundary value formulation in which the fields are determined successively along each null hypersurface. Second, it allows one to formulate different versions of the characteristic problem, depending on the nature of the boundary. We discuss the asymptotic initial value problem at large affine parameter, the local vertex problem at the central geodesic, and the null-boundary problem associated with two intersecting null hypersurfaces. In the asymptotic case we identify the scalar monopole, the scalar news, the Bondi mass and the charge {aspect}, and derive the corresponding balance laws. As a non-trivial check, we analyze the scalar-free sector of the hierarchy. In this limit the system reduces to the $n$-dimensional Einstein-Maxwell equations, and the hierarchy must reproduce the Reissner-Nordstr\"om-Tangherlini black hole \cite{Reissner1916,Nordstrom1918,Tangherlini1963}. We show that this is indeed the case. {The relevance of this check goes beyond reproducing a known static solution. In the four-dimensional Schwarzschild limit, the corresponding affine-null coordinates reduce to Israel's regular representation of the extended Schwarzschild geometry, which provides an explicit coordinate system covering the four regions of the maximally extended Kruskal spacetime \cite{Israel1966}. Thus, the affine-null gauge is not merely a local horizon-penetrating coordinate choice: in the simplest black-hole case it is tied to a global regular description of the extended spacetime.}  Both the non-extremal and extremal branches are obtained directly from the affine-null equations. The non-extremal solution gives the charged generalization of the Israel-type affine-null coordinates \cite{Israel1966}, while the extremal branch must be derived separately rather than obtained as a naive zero surface gravity limit of the non-extremal solution. In both cases, the affine-null fields are related explicitly to the standard Eddington-Finkelstein representation \cite{Eddington1924,Finkelstein1958}.

The formulation developed here also provides a useful starting point for future applications. In particular, it is naturally suited to the study of charged scalar collapse in higher dimensions, including possible higher-dimensional analogues of Choptuik scaling and critical phenomena \cite{Choptuik1993,Gundlach2007,SorkinOren2005}. Since the hypersurface equations remain regular at surfaces where the outgoing expansion vanishes, the framework is also appropriate for tracking apparent horizons and for following the scalar field into the black-hole interior \cite{MadlerBaakeHosseiniWinicour2024}.

{The relevant point for these applications is not only that the formulation is characteristic, but that the affine-null gauge, together with the auxiliary variables, produces a genuinely sequential radial hierarchy. In many double-null or more general characteristic approaches, the metric and matter variables remain coupled through a system of radial-temporal partial differential equations. By contrast, the present hierarchy determines the fields successively along each null generator by ordinary radial integration, while the scalar field is advanced by a first-order transport equation. This provides a concrete practical advantage for collapse studies, where one must repeatedly tune initial data, resolve small near-critical scales, locate the first trapped surfaces and extract black-hole observables. The advantage is therefore not merely formal: the radial part of the characteristic problem is organized as an ordinary-differential hierarchy rather than as a coupled characteristic PDE system.}

{This feature also clarifies the relation with other characteristic approaches. The present affine-null construction has the distinguishing feature that the radial coordinate is not the areal radius, while the areal radius is instead evolved as a field. This allows the same hierarchy to remain adapted to regions where the areal radius ceases to be a good radial coordinate, while retaining a direct connection with asymptotic quantities at future null infinity. Thus the formalism is designed to connect, within one characteristic setting, the radiative regime at $\mathscr I^+$, the formation of marginally trapped surfaces, and the subsequent evolution into the black-hole interior.}

{The higher-dimensional extension is physically motivated by the fact that critical collapse and extremality are not dimension-independent in a trivial way. This is already clear in neutral scalar collapse, where the critical exponent, echoing period and other diagnostics depend on the spacetime dimension. With a charged scalar field, the Maxwell sector, the gauge coupling and the Reissner-Nordstr\"om-Tangherlini extremality condition introduce further $n$-dependent structure. Hence the role of charge near criticality cannot be inferred by a direct extrapolation from four dimensions. Thus, the role played by the charged sector near the threshold of black-hole formation is a genuine physical question, rather than a straightforward extrapolation of the four-dimensional problem. Therefore, in a charged collapse problem, the relevant question is not only whether the black-hole mass exhibits critical scaling, but also how the electric charge scales, how it affects the critical solution, and how the final state is positioned relative to the dimension-dependent extremal bound.}

{This issue has been analyzed in detail in four spacetime dimensions. For the Einstein-Maxwell system coupled to a charged complex scalar field, the four-dimensional analyses of charged critical collapse established scaling laws and universality properties for both the black-hole mass and charge \cite{GundlachMartinGarcia1996,HodPiran1997}.
The dimensional dependence is already known to be important in the neutral scalar problem. Higher-dimensional scalar collapse has been studied through self-similar solutions, Choptuik scaling in six dimensions, numerical evolutions for $4\leq n\leq 11$, analyses of the dimension dependence of the critical exponent, and evolutions in Painlev\'e--Gullstrand coordinates \cite{Frolov1998,GarfinkleCutlerDuncan1999,SorkinOren2005,BlandPrestonBeckerKunstatterHusain2005,TavesKunstatter2011}. These works show that the critical exponent, the echoing period and other geometric diagnostics depend on the spacetime dimension. More recent analyses treating the dimension as a continuous parameter further support the view that $n$ controls part of the structure of the critical solution \cite{EckerEckerGrumillerJechtl2026,EckerEckerGrumiller2026}. By contrast, the corresponding higher-dimensional charged scalar problem is much less developed. Charged scalar collapse has been considered in de Sitter backgrounds in several dimensions \cite{ZhangZhangZouWang2015}, but a systematic asymptotically flat $n$-dimensional analogue of the four-dimensional charged scalar critical-collapse problem is still missing.}

A further motivation for the present formulation comes from recent developments concerning the third law of black-hole mechanics. The classical expectation that an exactly extremal black hole cannot form in finite time goes back to the laws of black-hole mechanics and to Israel's formulation of the third law \cite{BardeenCarterHawking1973,Israel1986ThirdLaw}. This expectation has now been shown not to hold, at least within certain matter models. In particular, Kehle and Unger constructed smooth spherically symmetric solutions of the Einstein-Maxwell-charged scalar field system which form an exactly extremal Reissner-Nordstr\"om black hole after a finite advanced time \cite{KehleUnger2025ThirdLaw}. They also showed, in the Einstein-Maxwell-Vlasov system, that extremal black holes can arise as critical solutions at the threshold between dispersion and collapse \cite{KehleUnger2024ExtremalCritical}. More recently, numerical evidence in five-dimensional vacuum gravity has shown that an extremal rotating black hole can form in finite time, suggesting that violations of the third law are not merely artifacts of charged matter models \cite{CrumpGadiouxReallSantos2026}. Recent work has also shown that higher-dimensional critical collapse contains additional geometric information beyond the usual mass-scaling exponent and echoing period. In particular, a new critical parameter was identified in Ref.~\cite{EckerEckerGrumiller2025}, associated with the angle formed, at the center of the critical spacetime, by the curves along which the null energy condition (NEC) is saturated. This ``NEC angle'' provides a geometric diagnostic of the critical solution and appears to encode universal information about the spacetime dimension. 

{These developments make the present hierarchy physically relevant in a concrete way. The system provides, in the same affine-null framework, the local charge, the quasi-local mass, the asymptotic Bondi mass and total charge, and the horizon condition associated with the vanishing of the outgoing expansion. These are precisely the quantities needed to study how charged collapse depends on spacetime dimension, whether a genuinely charged critical solution can arise, and how the approach to extremality depends on dimension and matter content.}

These results indicate that characteristic formulations are especially well suited to this type of questions. A hierarchical characteristic system for the $n$-dimensional Einstein equations, such as the one developed here in spherical symmetry, provides a natural framework for studying the approach to extremality, the formation of null horizons, and the role played by dimension, charge and matter content in possible violations of the third law.

{Therefore, the purpose of the $n$-dimensional construction is not merely to make the affine-null hierarchy as general as possible. Rather, it is to provide a characteristic framework with three specific advantages: a sequential radial ordinary-differential hierarchy, a coordinate choice adapted both to the asymptotic region along null directions and to black-hole interiors, and direct access to the geometric observables controlling charged collapse and extremality.}

The organization of this work is as follows. In Sec.~II we derive the $n$-dimensional Einstein-Maxwell-scalar equations in spherical symmetry and affine-null coordinates. In Sec.~III we introduce the auxiliary variables and cast the main equations into hierarchical form. We then discuss the corresponding characteristic initial-boundary value problems: the asymptotic problem, the local vertex problem and the null-boundary formulation. In Sec.~IV we use the scalar-free sector to recover the Reissner-Nordstr\"om-Tangherlini solution, including both non-extremal and extremal branches. We conclude with a summary and a discussion of possible extensions.

\section{Field equations for the Einstein-Maxwell-scalar system}

We begin with the Einstein-Hilbert action coupled to an electromagnetic field and a charged complex scalar field in $n$ dimensions,
\begin{equation}
    S=\int d^nx\sqrt{-g} \left[ \frac{g^{ab}R_{ab}}{2\kappa} - \frac{F_{ab}F^{ab}}{4\Omega_{n-2}} - \frac{1}{2} \left((\overline{D^a\Phi})(D_a\Phi) \right) \right] .
    \label{eq:action}
\end{equation}
Here $g_{ab}$ is an $n$-dimensional Lorentzian metric ($n\ge3$), $\nabla_a$ is its associated covariant derivative, $R_{ab}$ is its Ricci tensor, $\kappa$ is the gravitational coupling, and $\Omega_{n-2}$ is the area of the unit $(n-2)$-sphere. The Faraday tensor is
\begin{equation}
    F_{ab}=\nabla_aA_b-\nabla_bA_a,
\end{equation}
and the gauge covariant derivative is
$D_a=\nabla_a+iqA_a.$

Variation of the action gives the trace-reversed Einstein equations
\begin{equation}
    E_{ab}:=R_{ab}-\kappa\left(T_{ab}-\frac{1}{n-2}g_{ab}T\right)=0,
    \qquad
    T:=g^{ab}T_{ab},
    \label{eq:Einstein-trace-reversed}
\end{equation}
where $T_{ab}=T_{ab}^{(EM)}+T_{ab}^{(\Phi)}$ and
\begin{equation}
\begin{split}
    T_{ab}^{(EM)}=&\frac{1}{\Omega_{n-2}}\left(F_{ac}F_b{}^c-\frac14g_{ab}F_{cd}F^{cd}\right), \\
    T_{ab}^{(\Phi)}=&
    \frac{1}{2} \left\{\overline{D_a\Phi}D_b\Phi+\overline{D_b\Phi}D_a\Phi
    -g_{ab}\left[g^{cd}(\overline{D_c\Phi})(D_d\Phi)\right]\right\} .
\end{split}
\label{eq:stress-energy}
\end{equation}
The Maxwell equations are
\begin{equation}
    E_b:=\nabla_aF^a{}_{b}+\Omega_{n-2} j_b=0,
    \label{eq:Maxwell}
\end{equation}
with scalar current
\begin{equation}
    j_a=\frac{iq}{2}\left(\overline\Phi D_a\Phi-\Phi\overline{D_a\Phi}\right)
    =-q\,\mathrm{Im}\left(\overline{\Phi}D_a\Phi\right);
    \label{eq:scalar-current}
\end{equation}
while the resulting scalar equation reads
\begin{equation}
    E:=D^aD_a\Phi=0 .
    \label{eq:KG-general}
\end{equation}
Integrating Eq.~\eqref{eq:Maxwell} over a spacelike volume gives the total charge enclosed by its boundary,
\begin{equation}
    Q = \frac{1}{\Omega_{n-2}}\int_{\partial\sigma}*F .
\end{equation}

We now impose spherical symmetry and an affine null representation of the metric, namely
\begin{equation}
    ds^2=-V(w,\lambda)dw^2+2\epsilon\,dw\,d\lambda+r^2(w,\lambda)q_{AB}dx^A dx^B,
    \label{eq:affine-null-metric}
\end{equation}
where $\epsilon^2=1$, with $\epsilon=-1$ for outgoing and $\epsilon=1$ for ingoing null hypersurfaces $w=$const. The non-vanishing components of the inverse metric are
\begin{equation}
    g^{w\lambda}=\epsilon,
    \qquad
    g^{\lambda\lambda}=V,
    \qquad
    g^{AB}=\frac{1}{r^2}q^{AB},
    \label{eq:inverse-metric}
\end{equation}
and $\sqrt{-g}=r^{n-2}\sqrt{q}$. A convenient null frame is
\begin{equation}
    \ell^a\partial_a=\partial_\lambda,
    \qquad
    n^a\partial_a=-\epsilon\partial_w-\frac{V}{2}\partial_\lambda,
    \qquad
    \ell^a n_a=-1 .
    \label{eq:null-frame}
\end{equation}
Due to spherical symmetry, the scalar field does not depend on the angular coordinates,  $\Phi=\Phi(w,\lambda)$. We choose without loss of generality the radial gauge
\begin{equation}
    A_\lambda=0,
    \qquad
    A_a dx^a=\alpha(w,\lambda)\,dw,
\end{equation}
so that the only non trivial component of the Faraday tensor reads $F_{w\lambda}=-F_{\lambda w}=-\alpha_{,\lambda}.$

Under these assumptions, the non-trivial components of the Einstein equations reduce to
\begin{equation}
\begin{aligned}
(E_{ww}):\quad 0={}&
\frac{1}{2r}
\left[
(n-2)\left(-2r_{,ww}+\epsilon\left(V_{,\lambda}r_{,w}-r_{,\lambda}V_{,w}\right)\right)
+
\frac{V}{r^{n-3}}
\left(V_{,\lambda}r^{n-2}\right)_{,\lambda}
\right]
\\
&-
\kappa
\left[
\frac{n-3}{n-2}
\frac{V F_{w\lambda}^{2}}{\Omega_{n-2}}
+
|D_w\Phi|^2
\right],
\\
(E_{w\lambda}):\quad 0={}&
-\frac{1}{2\epsilon r}
\left[
2(n-2)\epsilon r_{,w\lambda}
+
\frac{1}{r^{n-3}}
\left(V_{,\lambda}r^{n-2}\right)_{,\lambda}
\right]
\\
&+
\kappa\epsilon\frac{n-3}{n-2}
\frac{F_{w\lambda}^{2}}{\Omega_{n-2}}
-
\kappa\,\mathrm{Re}\left(\overline{D_w\Phi}D_\lambda\Phi\right),
\\
(E_{\lambda\lambda}):\quad 0={}&
\frac{n-2}{r}r_{,\lambda\lambda}
+
\kappa|D_\lambda\Phi|^2,
\\
(q^{AB}E_{AB}):\quad 0={}&
(n-2)
\left[(n-3)-\frac{1}{r^{n-4}}
\left[r^{n-3}\left(r_{,\lambda}V+2\epsilon r_{,w}\right)\right]_{,\lambda}
\right]
-
\kappa\frac{r^2F_{w\lambda}^{2}}{\Omega_{n-2}} .
\end{aligned}
\label{eq:Einstein-components}
\end{equation}
while the Maxwell equations become
\begin{equation}
\begin{aligned}
(E_w):\quad 0={}&
\epsilon\left(r^{n-2}\alpha_{,\lambda}\right)_{,w}
+V\left(r^{n-2}\alpha_{,\lambda}\right)_{,\lambda}
-\Omega_{n-2}q r^{n-2}\,
\mathrm{Im}\left(\overline{\Phi}D_w\Phi\right),
\\
(E_\lambda):\quad 0={}&
\epsilon\left(r^{n-2}\alpha_{,\lambda}\right)_{,\lambda}
+\Omega_{n-2}q r^{n-2}\,
\mathrm{Im}\left(\overline{\Phi}D_\lambda\Phi\right).
\end{aligned}
\label{eq:Maxwell-components}
\end{equation}
and the massless scalar wave equation reads
\begin{equation}
    (E):\quad 0 =D_w\left(r^{n-2}D_\lambda\Phi\right)
    +D_\lambda\left(r^{n-2}D_w\Phi\right)
    +\epsilon D_\lambda\left(r^{n-2}V D_\lambda\Phi\right).
    \label{eq:scalar-wave-massless}
\end{equation}
Moreover, the charge $Q$ relates to $\alpha$ as:
\begin{equation}
    Q=-\epsilon r^{n-2}\alpha_{,\lambda}.
    \label{eq:charge-def}
\end{equation}

\section{Characteristic initial-boundary value problem}

The formulation of the characteristic initial-boundary value problem (CIBVP) for {a spherically symmetric charged scalar field} is based on the {twice} contracted Bianchi identities, together with {its} corresponding {electomagnetic analogue}. Once a suitable subset of the Einstein-Maxwell-scalar {field} equations is imposed throughout the spacetime domain, the remaining equations are not independent: they either follow identically from the main system or reduce to equations that must be imposed at the boundary $\lambda=\lambda_{\mathcal B}$. We refer to the former as trivial equations and to the latter as supplementary equations.

For the affine-null system considered here, we choose as main equations
\begin{equation}
    E_{\lambda\lambda}=0,
    \qquad
    q^{AB}E_{AB}=0,
    \qquad
    E_\lambda=0,
    \qquad
    E=0.
    \label{eq:main-equations}
\end{equation}
{The equations \eqref{eq:main-equations} give rise to a hypersurface-evolution algorithm on the computational domain of interest that is foliated with a family of null hypersurfaces $w=$constant.}

The supplementary equations are
\begin{equation}
    E_{ww}=0,
    \qquad
    E_w=0,
    \label{eq:supplementary-equations}
\end{equation}
and are imposed at the boundary, $\mathcal B$ {of the domain}. They provide the boundary evolution equations for the data specified on $\mathcal B$ and serve as consistency checks during an evolution. The mixed Einstein equation
\begin{equation}
    E_{w\lambda}=0,
    \label{eq:trivial-equation}
\end{equation}
is then a trivial equation in the sense that it is propagated by the main system once the supplementary equations hold at the boundary. Equivalently, one may interchange the roles of $q^{AB}E_{AB}$ and $E_{w\lambda}$, using $E_{w\lambda}$ as a main equation and treating the angular trace equation as the trivial one \cite{Tamburino:1966zz}.

A central goal of the characteristic formulation is to cast the main field equations into a hierarchy of hypersurface equations. In Bondi-Sachs coordinates this structure arises naturally: once the radiative data are specified on an initial null hypersurface, the remaining metric variables are obtained sequentially by radial integration, and the wave equation updates the dynamical field to the next null slice \cite{Madler:2016xju}. In affine-null coordinates, however, this hierarchy is not immediate. The angular Einstein equation contains the derivative $r_{,w}$, which prevents direct radial integration in terms of the original variables. 
This obstruction is a consequence of using the affine parameter $\lambda$ instead of the areal radius as radial coordinate. As was shown in four dimensions in Ref.~\cite{Madler:2025ibn}, the introduction of suitable auxiliary variables restores the hierarchical structure. We now generalize this construction to $n$-dimensions.

We introduce the auxiliary variables:
\begin{eqnarray}
    Z &=& r^{n-3}\left(r_{,\lambda}V+2\epsilon r_{,w}\right),
    \label{eq:Zn}\
    \\
    \mathcal{L} &=& \left[2r\left(\Phi_{,w}r_{,\lambda}-r_{,w}\Phi_{,\lambda}\right)
    +\frac{\epsilon Z\Phi_{,\lambda}}{r^{n-4}}\right]\frac{1}{r_{,\lambda}}.
    \label{eq:L-def}
\end{eqnarray}
The main equations are then cast into the hierarchy
\begin{eqnarray}
    r_{,\lambda\lambda}
    &=&
    -\frac{\kappa r}{n-2}|\Phi_{,\lambda}|^2,
    \label{eq:hipr}
    \\
    Q_{,\lambda}
    &=&
    \Omega_{n-2}q r^{n-2}\operatorname{Im}\left(\overline{\Phi}\Phi_{,\lambda}\right),
    \label{eq:charg}
    \\
    \alpha_{,\lambda}
    &=&
    -\frac{\epsilon Q}{r^{n-2}},
    \label{eq:alpharef}
    \\
    Z_{,\lambda}
    &=&
    (n-3)r^{n-4}
    -\frac{\kappa}{n-2}\frac{Q^2}{\Omega_{n-2}r^{n-2}},
    \label{eq:zfre}
    \\
    \mathcal{L}_{,\lambda}
    &=&
    -\frac{n-4}{2r}r_{,\lambda}\mathcal{L}
    -\frac{n-2}{2r^{n-3}}\epsilon Z\Phi_{,\lambda}
    +iq\epsilon\Phi\frac{Q}{r^{n-3}}
    -2iq\alpha(r\Phi)_{,\lambda}
    -(n-4)iq\alpha r_{,\lambda}\Phi,
    \label{eq:L-hier}
    \\
    V_{,\lambda\lambda}
    &=&
    -\frac{(n-2)(n-3)}{r^2}
    \left(1-\frac{Zr_{,\lambda}}{r^{n-3}}\right)
    +\frac{\kappa}{\Omega_{n-2}}\frac{3n-8}{n-2}\frac{Q^2}{r^{2n-4}}
    -\frac{\kappa\epsilon}{r}\operatorname{Re}\left(\overline{\Phi_{,\lambda}}\mathcal{L}\right)
    -2\epsilon\kappa q\alpha\operatorname{Im}\left(\overline{\Phi}\Phi_{,\lambda}\right).
    \label{eq:V-hier}
\end{eqnarray}
The scalar evolution equation follows from Eq.~\eqref{eq:L-def},
\begin{equation}
    \Phi_{,w}=\frac{\mathcal{L}}{2r}-\frac{\epsilon V\Phi_{,\lambda}}{2}.
    \label{eq:Phi-evolution}
\end{equation}
Thus $r$, $Q$, $\alpha$, $Z$, $\mathcal{L}$ and $V$ can be solved successively along each null hypersurface, while $\Phi$ is evolved through Eq.~\eqref{eq:Phi-evolution}.
The evolution is determined once appropriate data are specified on a initial null hypersurface $\mathcal{N}_0$ defined by $w_0=$constant, and on the boundary $\lambda=\lambda_{\mathcal{B}}$. As discussed in \cite{Madler:2025ibn}, this data consist in given $\Phi(0,\lambda)$ at the given null hypersurface $\mathcal{N}_0$ and the fields 
\begin{equation}
r(w,\lambda_{\mathcal{B}}),\qquad
r_{,\lambda}(w,\lambda_{\mathcal{B}}),\qquad
Q(w,\lambda_{\mathcal{B}}),\qquad
\alpha(w,\lambda_{\mathcal{B}}),\qquad
\mathcal{L}(w,\lambda_{\mathcal{B}}),\qquad
V(w,\lambda_{\mathcal{B}}),\qquad
V_{,\lambda}(w,\lambda_{\mathcal{B}}),
\end{equation}
on the boundary $\mathcal{B}$.
In what follows, we shall consider different boundary configurations and their associated initial data.

\subsection{Asymptotic initial value problem}

We now turn to the characteristic initial value problem in the asymptotic
region, namely for large values of the affine parameter $\lambda$ along each
null hypersurface $w=\mathrm{constant}$. The purpose of this analysis is to
identify the free radiative data at null infinity and to relate the integration
functions appearing in the affine null hierarchy to invariant quantities, such
as the Bondi mass and the total charge.

Throughout this subsection we assume $n>3$ and introduce
\begin{equation}
    d=n-2,\qquad
    p=n-3=d-1,\qquad
    \sigma=\frac{d}{2}=\frac{n-2}{2}.
\end{equation}

In the following we will also make use of the $n$-dimensional Misner-Sharp mass  defined by \cite{Maeda:2007uu}
\begin{equation}
{g^{rr} = 1- \frac{\kappa }{(n-2)\Omega_{n-2}}\frac{2 M}{r^{n-3}}\;\;\Rightarrow\;\;}
    M = \frac{(n-2)\Omega_{n-2}}{2\kappa} r^{n-3}\left(1-g^{ab}r_{,a}r_{,b}\right),
    \label{eq:MS-general}
\end{equation}
which in the affine null coordinates reduces to,
\begin{equation}
    M=\frac{(n-2)\Omega_{n-2}}{2\kappa}r^{n-3}
    \left(1-2\epsilon r_{,w}r_{,\lambda}-V r_{,\lambda}^2\right).
    \label{eq:MS-affine}
\end{equation}
{The Bondi mass is defined from  the asymptotic limit of the Misner-Sharp mass,}
\begin{equation}
    m_B=\lim_{\substack{\lambda\to\infty\\ w=\mathrm{const}}}M .
    \label{eq:Bondi-general}
\end{equation}

 We assume that the fields admit regular asymptotic expansions on each null hypersurface. For the massless charged scalar field, the natural radiation condition in $n$ spacetime dimensions is that $\lambda^\sigma\Phi$ have a finite limit as $\lambda\to\infty$, with $\sigma=(n-2)/2$. This is the standard radiative order for massless scalar waves in $n$ dimensions. The minimal electromagnetic coupling does not change this leading power; rather, it replaces the ordinary derivative of the scalar monopole by its gauge-covariant counterpart. With this fall-off, the scalar charge current has the finite-flux asymptotic order used in the Maxwell analysis of Ref.~\cite{SatishchandranWald2019}. Accordingly, we write
\begin{equation}
    \Phi(w,\lambda)
    =\frac{\Phi_{[1]}(w)}{\lambda^\sigma}
    +\frac{\Phi_{[2]}(w)}{\lambda^{\sigma+1}}+O(\lambda^{-\sigma-2}).
    \label{eq:asymp-Phi}
\end{equation}

The areal radius is expanded consistently with asymptotic flatness as
\begin{equation}
    r(w,\lambda)
    =H(w)\lambda+R_\infty(w)+\frac{r_{[-p]}(w)}{\lambda^p}+O(\lambda^{-p-1}).
    \label{eq:asymp-r-ansatz}
\end{equation}
Here $H(w)$ measures the asymptotic normalization of the affine parameter,
while $R_\infty(w)$ is the leading asymptotic shift of the areal coordinate.

The first equation of the hierarchy, Eq.~\eqref{eq:hipr}, fixes the first
non-trivial correction to the areal radius. Substituting the scalar expansion
\eqref{eq:asymp-Phi} into Eq.~\eqref{eq:hipr} gives
\begin{equation}
    r_{[-p]}
    =-\frac{\kappa H}{4p}|\Phi_{[1]}|^2.
\end{equation}
Thus
\begin{equation}
    r(w,\lambda)
    =H\lambda+R_\infty-\frac{\kappa H}{4p}
    \frac{|\Phi_{[1]}|^2}{\lambda^p}+
    O(\lambda^{-p-1}).
    \label{eq:asymp-r}
\end{equation}

The charge equation \eqref{eq:charg} then determines the asymptotic behavior
of the charge function. Since the leading scalar contribution to the current
appears at order $\lambda^{-d-2}$, one obtains
\begin{equation}
    Q(w,\lambda)
    =Q_\infty(w)+\frac{\Omega_{n-2}qH^d
    \operatorname{Im}\left(\overline{\Phi}_{[1]}\Phi_{[2]}\right)}
    {\lambda}+O(\lambda^{-2}).
    \label{eq:asymp-Q}
\end{equation}
The integration function $Q_\infty(w)$ is the total charge measured at null
infinity.

Once $Q$ is known, the Maxwell potential is obtained directly from
Eq.~\eqref{eq:alpharef}. This gives
\begin{equation}
\alpha(w,\lambda)=\alpha_\infty(w)+\frac{\epsilon Q_\infty(w)}{pH^d\lambda^p}+O(\lambda^{-p-1}).
    \label{eq:asymp-alpha}
\end{equation}
The function $\alpha_\infty(w)$ is the residual asymptotic gauge potential.

The next hypersurface equation determines $Z$. Using
Eqs.~\eqref{eq:asymp-r} and \eqref{eq:asymp-Q} in Eq.~\eqref{eq:zfre}, one
finds
\begin{equation}
    Z(w,\lambda)=\frac{(H\lambda+R_{\infty})^p}{H}+Z_\infty(w)+O(\lambda^{-1}).
    \label{eq:asymp-Z}
\end{equation}
The integration function $Z_\infty(w)$ will be identified below with the
Bondi mass aspect.

The metric coefficient $V$ is most efficiently obtained from the definition
of $Z$, Eq.~\eqref{eq:Zn}. Substituting the asymptotic expansions of $r$ and
$Z$ gives
\begin{equation}
    V(w,\lambda)
    =-2\epsilon\,\frac{H_{,w}}{H}\,\lambda+\frac{1}{H^2}\left(
1-2\epsilon H R_{\infty,w}
\right)+\frac{1}{H r^p}\left[Z_\infty
+\frac{\epsilon\kappa}{2p}\left(H^d |\Phi_{[1]}|^2\right)_{,w}\right]+O(\lambda^{-p-1}).
    \label{eq:asymp-V}
\end{equation}
Thus the leading term of $V$ is controlled by the variation of the
asymptotic normalization $H(w)$, while the finite part contains the
asymptotic shift $R_\infty(w)$.

The leading behavior of the auxiliary field $\mathcal{L}$ follows either from the
transport equation \eqref{eq:Phi-evolution} or directly from its definition.
At leading order one obtains
\begin{equation}
    \mathcal{L}(w,\lambda)=2H^{1-\sigma}
    \frac{d}{dw}\left(H^\sigma\Phi_{[1]}\right)\lambda^{1-\sigma}+O(\lambda^{-\sigma}).
    \label{eq:asymp-L}
\end{equation}
This coefficient contains the time derivative of the leading scalar
monopole, and therefore it is the affine null precursor of the scalar news.

We now identify the mass aspect. From the Misner-Sharp mass
\eqref{eq:MS-affine} and the definition of $Z$, Eq.~\eqref{eq:Zn}, one has
\begin{equation}
    M=\frac{d\Omega_{n-2}}{2\kappa}\left(r^p-r_{,\lambda}Z\right).
\end{equation}
Using the asymptotic expansions \eqref{eq:asymp-r} and \eqref{eq:asymp-Z}, the resulting Bondi mass reads
\begin{equation}
m_B:=\lim_{\lambda\to\infty}M
    =-\frac{d\Omega_{n-2}}{2\kappa}H Z_\infty.
    \label{eq:Bondi-mass-asymp}
\end{equation}

To make contact with the standard inertial description at null infinity, we
introduce the asymptotic areal coordinate
\begin{equation}
    x(w,\lambda)=H(w)\lambda+R_\infty(w)+O(\lambda^{-p})
    \label{eq:asymp-x}
\end{equation}
and define the Bondi time $w_B$ by
\begin{equation}
    dw_B=\frac{dw}{H(w)},\qquad\frac{dw_B}{dw}=\frac{1}{H}.
    \label{eq:Bondi-time}
\end{equation}
In these variables the metric approaches
\begin{equation}
    ds^2=-dw_B^2+2\epsilon\,dw_B\,dx+x^2q_{AB}dx^A dx^B+O(x^{-1}),
    \label{eq:Bondi-frame-metric}
\end{equation}
{which also justifies calling $H$ a redshift factor. Indeed in case of $\epsilon=-1$, the limit $H\rightarrow0$ for $w=w_H>0$ indicates the formation time of a future event horizon at $w=w_H$.}
The frame in which $H=1$ is the Bondi frame. In this frame the asymptotic
fields take the simpler form
\begin{equation}
    r=\lambda+R_\infty-\frac{\kappa}{4p}\frac{|\Phi_{[1]}|^2}{\lambda^p}+O(\lambda^{-p-1}),
\end{equation}
and
\begin{equation}
    V=1-2\epsilon R_{\infty,w_B}-\frac{2\kappa m_B}{d\Omega_{n-2}r^p}+
\frac{\epsilon\kappa}{2p\,r^p}
\frac{d|\Phi_{[1]}|^2}{dw_B}+O(r^{-p-1}) .
\end{equation}

Now let us discuss the balance laws. To this end, we first define the scalar monopole measured at null infinity by
\begin{equation}
    C(w_B):=\lim_{\lambda\to\infty}(H\lambda)^\sigma\Phi=H^\sigma\Phi_{[1]}.
    \label{eq:scalar-monopole}
\end{equation}
The next coefficient in the Bondi expansion,
\begin{equation}
    \Phi=\frac{C(w_B)}{r^\sigma}+\frac{C_{[2]}(w_B)}{r^{\sigma+1}}+O(r^{-\sigma-2}),
\end{equation}
is related to the affine null coefficients by
\begin{equation}
    C_{[2]}=H^{\sigma+1}\Phi_{[2]}+\sigma R_\infty C.
    \label{eq:scalar-subleading}
\end{equation}
For $n=4$, one has $\sigma=1$, and this relation reduces to the familiar
connection between the Newman Penrose coefficient and the affine expansion.

The scalar news is defined as the Bondi time derivative of the scalar
monopole,
\begin{equation}
    N(w_B):=\frac{dC}{dw_B}.
    \label{eq:scalar-news}
\end{equation}
Since $dw_B=dw/H$, this is equivalently
\begin{equation}\label{news2}
    N=H\frac{d}{dw}\left(H^\sigma\Phi_{[1]}\right).
\end{equation}
{We further note that the news function $N$ can be extracted from the auxiliary hypersurface field $\mathcal{L}$ at large  values of $\lambda$, compare  \eqref{eq:asymp-L} with \eqref{news2}}
The asymptotic Maxwell potential combines with the scalar news into the
gauge covariant quantity
\begin{equation}
    \mathcal N:=N+iqH\alpha_\infty C.
    \label{eq:gauge-covariant-news}
\end{equation}
It is this combination, rather than $N$ alone, that enters the asymptotic
fluxes before fixing the residual electromagnetic gauge.

Evaluating the supplementary equations at null infinity gives the balance
laws
\begin{equation}
    \frac{dm_B}{dw_B}=\epsilon\Omega_{n-2}|\mathcal N|^2,
    \label{eq:Bondi-mass-balance}
\end{equation}
and
\begin{equation}
    \frac{dQ_\infty}{dw_B}=-\Omega_{n-2}q\operatorname{Im}\left(\overline{C}\mathcal N\right).
    \label{eq:charge-balance}
\end{equation}
Equivalently,
\begin{equation}
    \frac{dQ_\infty}{dw_B}=-\Omega_{n-2}q\left[\operatorname{Im}\left(\overline{C}N\right)+qH\alpha_\infty |C|^2\right].
\end{equation}
The remaining electromagnetic gauge freedom may be used to set $\alpha_\infty=0$. In this gauge $\mathcal N=N$, and the balance laws reduce
to
\begin{equation}
    \frac{dm_B}{dw_B}=\epsilon\Omega_{n-2}|N|^2,
    \qquad \frac{dQ_\infty}{dw_B}=-\Omega_{n-2}q\operatorname{Im}\left(\overline{C}N\right).
    \label{eq:balance-alpha-zero}
\end{equation}
For outgoing null hypersurfaces, $\epsilon=-1$, the first equation gives the
expected decrease of the Bondi mass whenever scalar radiation is present.

The asymptotic initial value problem can now be stated in intrinsic Bondi
quantities. In the gauge $\alpha_\infty=0$, one specifies the scalar news
$N(w_B)$ together with the initial values
\begin{equation}
    C(0), \qquad m_B(0), \qquad Q_\infty(0).
\end{equation}
The asymptotic evolution is then governed by the ordinary differential system
\begin{align}
    \frac{dC}{dw_B}&=N(w_B),
    \\
    \frac{dm_B}{dw_B}&=\epsilon\Omega_{n-2}|N(w_B)|^2,
    \\
    \frac{dQ_\infty}{dw_B}&=-\Omega_{n-2}q
    \operatorname{Im}\left(\overline{C}(w_B)N(w_B)\right).
\end{align}
Thus the free function $N(w_B)$ determines the time dependence of the scalar
monopole, while the balance laws determine the corresponding evolution of
the Bondi mass and the total charge. In the uncharged case $q=0$, the charge
is constant and the asymptotic data reduce to the scalar monopole, the Bondi
mass and the scalar news.

\subsection{Local initial value problem at the vertex}

We now consider the local characteristic initial value problem in a neighborhood of the central geodesic of spherical symmetry. Let $c(w)$ denote this central timelike geodesic. The null hypersurfaces $w=\mathrm{constant}$ are assumed to emanate from $c(w)$, and the origin of the affine parameter is chosen at the vertex,
\begin{equation}
    \lambda=0
    \qquad\Longleftrightarrow\qquad
    r=0 .
\end{equation}
The affine parameter is normalized by requiring
\begin{equation}
    r(w,0)=0,
    \qquad
    r_{,\lambda}(w,0)=1 .
\end{equation}
Regularity of the metric at the vertex then fixes the leading behavior of the remaining metric function,
\begin{equation}
    V(w,0)=1,
    \qquad
    V_{,\lambda}(w,0)=0 .
\end{equation}
Therefore, in a sufficiently small neighborhood of the vertex, the line element has the local flat-space form
\begin{equation}
    ds^2=-\left[1+O(\lambda^2)\right]dw^2+2\epsilon\,dw\,d\lambda+\lambda^2q_{AB}dx^A dx^B+O(\lambda^3).
\end{equation}
The electromagnetic gauge potential has vertex value $\alpha(w,0)=\alpha_0(w)$, which is pure gauge. Regularity of the charge function also requires $Q(w,0)=0$. Moreover, since $r(w,0)=0$ for every value of $w$, differentiating this condition along the central geodesic gives $r_{,w}(w,0)=0.$
Thus, near $\lambda=0$,
\begin{equation}
    r=\lambda+O(\lambda^3),
    \qquad
    r_{,\lambda}=1+O(\lambda^2),
    \qquad
    V=1+O(\lambda^2),
    \qquad
    r_{,w}=O(\lambda^3).
\end{equation}
These regularity conditions immediately determine the leading behavior of the auxiliary variables. In particular, Eq.~\eqref{eq:Zn} gives
\begin{equation}
    Z=\lambda^{n-3}+O(\lambda^{n-1}),
\end{equation}
so that $Z(w,0)=0$ for $n>3$. Similarly, from Eq.~\eqref{eq:L-def}, assuming the regular Taylor expansion
\begin{equation}
    \Phi(w,\lambda)=\Phi_{(0)}(w)+\Phi_{(1)}(w)\lambda+\frac{1}{2}\Phi_{(2)}(w)\lambda^2+O(\lambda^3),
    \label{eq:vertex-Phi-expansion}
\end{equation}
one finds $\mathcal{L}=O(\lambda)$ and therefore $\mathcal{L}(w,0)=0$.

It is useful to introduce a notation for the scalar contribution to the radial charge current that reduces directly to the four-dimensional convention used in Ref.~\cite{Madler:2025ibn}. We define
\begin{equation}
    \mathcal J:=i\left(\overline{\Phi}\Phi_{,\lambda}-\Phi\overline{\Phi}_{,\lambda}\right)=-2\operatorname{Im}\left(\overline{\Phi}\Phi_{,\lambda}\right).
\end{equation}
The Taylor expansion of this quantity at the vertex is
\begin{equation}
    \mathcal J=\mathcal J_{(0)}+\mathcal J_{(1)}\lambda+O(\lambda^2),
\end{equation}
where
\begin{equation}
    \mathcal J_{(0)}=i\left(\overline{\Phi}_{(0)}\Phi_{(1)}-\Phi_{(0)}\overline{\Phi}_{(1)}\right),
    \qquad
    \mathcal J_{(1)}=i\left(\overline{\Phi}_{(0)}\Phi_{(2)}-\Phi_{(0)}\overline{\Phi}_{(2)}\right).
    \label{eq:vertex-J-coefficients}
\end{equation}
With this convention, Eq.~\eqref{eq:charg} becomes
\begin{equation}
    Q_{,\lambda}=-\frac{\Omega_{n-2}q}{2}r^{n-2}\mathcal J .
\end{equation}

We can now solve the hypersurface equations order by order in $\lambda$. The result is a local expansion determined entirely by the Taylor coefficients of the scalar field at the vertex, together with the gauge function $\alpha_0(w)$. Collecting the local expansions obtained from the hypersurface equations, one finds
\begin{align}
    r(w,\lambda)
    &=
    \lambda-\frac{\kappa}{6(n-2)}|\Phi_{(1)}|^2\lambda^3-\frac{\kappa}{12(n-2)}
\left(\overline{\Phi}_{(1)}\Phi_{(2)}+\Phi_{(1)}\overline{\Phi}_{(2)}\right)\lambda^4
    +O(\lambda^5),
    \label{eq:vertex-r-expansion}
    \\
    Q(w,\lambda)
    &=
    -\frac{\Omega_{n-2}q}{2}\left[\frac{\mathcal J_{(0)}}{n-1}\lambda^{n-1}+\frac{\mathcal J_{(1)}}{n}\lambda^n\right]
    +O(\lambda^{n+1}),
    \label{eq:vertex-Q-expansion}
    \\
    \alpha(w,\lambda)
    &=\alpha_0(w)+\frac{\epsilon\Omega_{n-2}q}{2}
    \left[\frac{\mathcal J_{(0)}}{2(n-1)}\lambda^2
    +\frac{\mathcal J_{(1)}}{3n}\lambda^3\right]
    +O(\lambda^4),
    \label{eq:vertex-alpha-expansion}
    \\
    Z(w,\lambda)
    &=
    \lambda^{n-3}
    -\frac{\kappa(n-3)(n-4)}{6(n-2)(n-1)}|\Phi_{(1)}|^2\lambda^{n-1}
    -\frac{\kappa(n-3)(n-4)}{12(n-2)n}
\left(\overline{\Phi}_{(1)}\Phi_{(2)}+\Phi_{(1)}\overline{\Phi}_{(2)}\right)\lambda^n
    \nonumber\\
    &\quad
    -\frac{\kappa\Omega_{n-2}q^2}{4(n-2)(n-1)^2(n+1)}\mathcal J_{(0)}^2\lambda^{n+1}
    -\frac{\kappa\Omega_{n-2}q^2}{2(n-2)n(n-1)(n+2)}\mathcal J_{(0)}\mathcal J_{(1)}\lambda^{n+2}
    +O(\lambda^{n+3}),
    \label{eq:vertex-Z-expansion}
    \\
    \mathcal{L}(w,\lambda)
    &=
    -\left(\epsilon\Phi_{(1)}+2iq\alpha_0\Phi_{(0)}\right)\lambda
    -\left[\epsilon\frac{n-2}{n}\Phi_{(2)}+2iq\alpha_0\Phi_{(1)}\right]\lambda^2
    +O(\lambda^3),
    \label{eq:vertex-L-expansion}
    \\
    V(w,\lambda)
    &=
    1+\frac{\kappa}{n-1}|\Phi_{(1)}|^2\lambda^2+O(\lambda^3),
    \label{eq:vertex-V-expansion}
    \\
    \Phi_{,w}
    &=
    -\epsilon\Phi_{(1)}-iq\alpha_0\Phi_{(0)}
    -\left[\epsilon\frac{n-1}{n}\Phi_{(2)}+iq\alpha_0\Phi_{(1)}\right]\lambda
    +O(\lambda^2).
    \label{eq:vertex-Phiw-expansion}
\end{align}
This expansion reduces to that presented in \cite{Madler:2025ibn}  for $n=4$. Note however, that  the auxiliary variable $Z$ used here is related to the one used in Ref.~\cite{Madler:2025ibn} by $Z_{\cite{Madler:2025ibn}}=Z-\lambda$.

The corresponding local vertex initial value problem can now be stated in a precise way. Let the initial scalar profile on $w=0$ admit a regular Taylor expansion at the vertex,
\begin{equation}
    \Phi(0,\lambda)=\sum_{k=0}^{\infty}\frac{\Phi_{(k)}(0)}{k!}\lambda^k .
\end{equation}
Given the gauge charge $q$ and the gauge function $\alpha_0(w)$ at the vertex, the hypersurface equations determine $r,Q,\alpha,Z,\mathcal{L},V$ order by order in $\lambda$. Once these fields are known, the transport equation determines the $w$-evolution of the Taylor coefficients of the scalar field. In particular,
\begin{equation}
    \Phi_{(0),w}=-\epsilon\Phi_{(1)}-iq\alpha_0\Phi_{(0)},
    \qquad
    \Phi_{(1),w}=-\epsilon\frac{n-1}{n}\Phi_{(2)}-iq\alpha_0\Phi_{(1)}.
\end{equation}
Thus the regularity of the scalar profile at the vertex provides a hierarchical local evolution system for all Taylor coefficients $\Phi_{(k)}(w)$.

\subsection{Null boundary value problem}
If the boundary is a null hypersurface, then we require data on the sphere $\Sigma_0=\mathcal N_0\cap\mathcal B$ and on the null hypersurfaces $\mathcal B$ and $\mathcal N_0$:
\begin{equation}
    {0<r|_{\Sigma_0}},
    \quad
    r_{,\lambda}|_{\Sigma_0},
    \quad
    r_{,w}|_{\Sigma_0},
    \quad
    Q|_{\Sigma_0},
    \quad
    \Phi|_{\Sigma_0},
    \quad
    \Phi_{,\lambda}|_{\mathcal N_0},
    \quad
    \Phi_{,w}|_{\mathcal B}.
\end{equation}
with $V(w,\lambda_{\mathcal{B}})=V_{,\lambda}(w,\lambda_{\mathcal{B}})=\alpha(w,\lambda_{\mathcal{B}})=0$. The condition $V(w,\lambda_{\mathcal B})=0$ ensures that the boundary $\mathcal B$ is a null hypersurface, while $V_{,\lambda}(w,\lambda_{\mathcal B})=0$ implies that its null generators are affinely parametrized by $w$. Without loss of generality, we can take $\lambda_{\mathcal{B}}=0$ and $\mathcal{N}_0$ placed at $w=0.$
{Under these conditions,  the supplementary equations evaluated at $\mathcal B$ give
\begin{equation}
\begin{aligned}
    r_{,ww}\Big|_{\mathcal B}
    &=
    -\frac{\kappa r}{n-2}|\Phi_{,w}|^2\Big|_{\mathcal B},
    \\
    Q_{,w}\Big|_{\mathcal B}
    &=
    -\Omega_{n-2}q r^{n-2}\mathrm{Im}\left(\overline\Phi\Phi_{,w}\right)\Big|_{\mathcal B},
    \\
    \left(r^{n-3}r_{,\lambda}\right)_{,w}\Big|_{\mathcal B}
    &=
    \frac{n-3}{2}\epsilon r^{n-4}
    -\frac{\kappa\epsilon}{2\Omega_{n-2}(n-2)}\frac{Q^2}{r^{n-2}}\Big|_{\mathcal B}.
\end{aligned}
\label{eq:boundary-evolution-massless}
\end{equation}}

\begin{figure}
    \centering
    \IfFileExists{datos_iniciales_scalar_field.png}{%
        \includegraphics[width=0.7\linewidth]{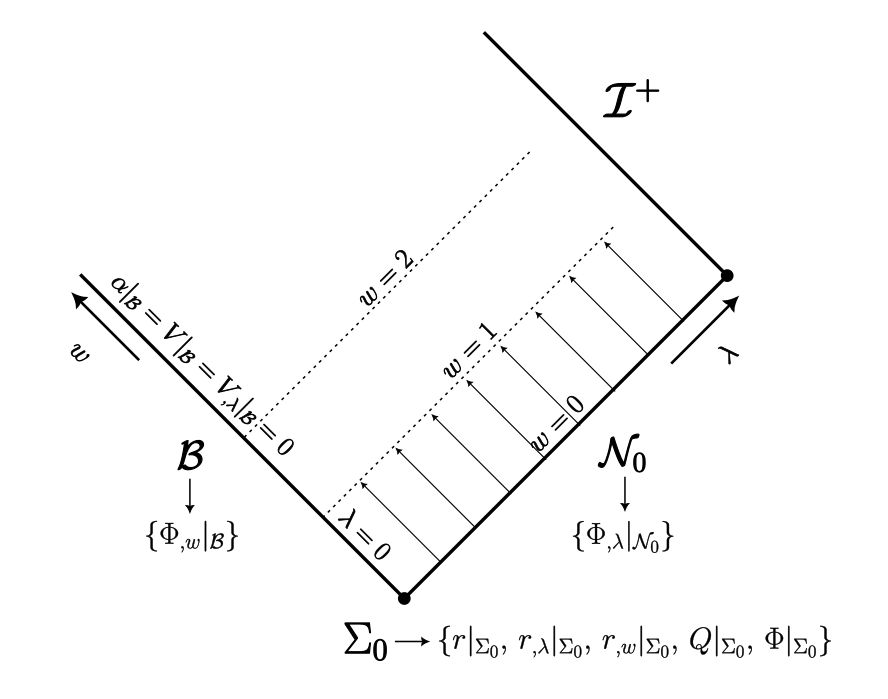}%
    }{%
        \fbox{\parbox{0.65\linewidth}{\centering datos\_iniciales\_scalar\_field.png}}%
    }
    \caption{\footnotesize{Initial data on the bifurcation sphere $\Sigma_0$ and on the null hypersurfaces $\mathcal B$ and $\mathcal N_0$ for the case $\epsilon=-1$.}}
    \label{scalar_field_IC}
\end{figure}

The null-boundary characteristic problem is evolved by combining boundary propagation with radial hypersurface integration. They are illustrated in Fig.~\ref{scalar_field_IC} for the case $\epsilon=-1$. First, the prescribed boundary data $N_{\mathcal B}(w)=\Phi_{,w}(w,0)$, together with the value $\Phi(0,0)$ at $\Sigma_0$, determines the scalar field on the null boundary $\mathcal B$. The supplementary equations on $\mathcal B$ then propagate the non-trivial boundary fields, in particular $r$, $Q$ and $r_{,\lambda}$, while the auxiliary fields $Z$ and $L$ are fixed there by their definitions. Independently, the initial scalar profile on the null hypersurface $\mathcal N_0$ is obtained from the data $F_\Phi(\lambda)=\Phi_{,\lambda}(0,\lambda)$ and the same value $\Phi(0,0)$ on $\Sigma_0$. Once these boundary and initial data are known, the hierarchical hypersurface equations determine $r$, $Q$, $\alpha$, $Z$, $L$ and $V$ along $\mathcal N_0$. The transport equation for the scalar field then gives $\Phi_{,w}$, allowing one to advance $\Phi$ to the next null hypersurface. The evolved scalar profile is used as the new initial data, and the same hierarchy is iterated to generate the characteristic evolution. We refer to \cite{Madler:2025ibn} for more details.

\section{Spherically symmetric electro vacuum black holes}

We now use the scalar-free sector as a non-trivial check of the affine-null hierarchy. When the scalar field is absent, the system reduces to the Einstein-Maxwell equations, and the resulting equations should reproduce the static, spherically symmetric Reissner-Nordström-Tangherlini black hole in $n$ dimensions~\cite{Tangherlini1963}. This provides a useful consistency test because the solution is known in standard coordinates, while here it will be obtained directly in the affine null coordinates. We address this problem in the context of a null boundary value problem.

We set $\Phi=0$. Consequently, $\mathcal{L}=0$ and $Q_{,\lambda}=0$. As before, let $p=n-3$. The relevant hierarchical equations are
\begin{align}
    r_{,\lambda\lambda}&=0,
    \label{eq:RN-hier-r}
    \\
    Q_{,\lambda}&=0,
    \label{eq:RN-hier-Q}
    \\
    \alpha_{,\lambda}&=-\frac{\epsilon Q}{r^{p+1}},
    \label{eq:RN-hier-alpha}
    \\
    Z_{,\lambda}&=p r^{p-1}-\frac{\kappa Q^2}{(p+1)\Omega_{n-2}}\frac{1}{r^{p+1}},
    \label{eq:RN-hier-Z}
    \\
    V_{,\lambda\lambda}&=-\frac{p(p+1)}{r^2}\left(1-\frac{Zr_{,\lambda}}{r^p}\right)
    +\frac{\kappa}{\Omega_{n-2}}\frac{3p+1}{p+1}\frac{Q^2}{r^{2p+2}}.
    \label{eq:RN-hier-V}
\end{align}
The corresponding boundary equations simplify to
\begin{eqnarray}
    r_{,ww}\Big|_{\mathcal B}&=&0,
    \label{eq:rwwn}
    \\
    Q_{,w}\Big|_{\mathcal B}&=&0,
    \label{eq:Qwnn}
    \\
    \left(r^p r_{,\lambda}\right)_{,w}\Big|_{\mathcal B}
    &=&
    \frac{p}{2}\epsilon r^{p-1}
    -\frac{\kappa\epsilon}{2\Omega_{n-2}(p+1)}\frac{Q^2}{r^{p+1}}.
    \label{eq:Hnn}
\end{eqnarray}
Thus, at the initial two-sphere $\Sigma_0$, the scalar-free problem is characterized by the following geometric and electromagnetic data:
\begin{equation}
    r|_{\Sigma_0},
    \qquad
    r_{,\lambda}|_{\Sigma_0},
    \qquad
    r_{,w}|_{\Sigma_0},
    \qquad
    Q|_{\Sigma_0}.
\end{equation}
Equations~\eqref{eq:RN-hier-Q} and \eqref{eq:Qwnn} imply that $Q$ is constant throughout the solution. In turn, Eq.~\eqref{eq:RN-hier-r} and \eqref{eq:rwwn} shows that the areal radius must be affine in $\lambda$, namely
\begin{equation}
    r(w,\lambda)=R(w)+H(w)\lambda,\qquad R(w)=r_0+r_Nw.
    \label{eq:Rdef}
\end{equation}
The fact that the coefficient proportional to $\lambda$ in $r$ is denoted by $H(w)$ is because, as we shall show later, it is in fact the redshift factor between a Bondi time $u$ and the coordinate time $w$ {($\epsilon=-1$).}
The remaining boundary equation then determines how the coefficient $H(w)=r_{,\lambda}(w,0)$ evolves along the boundary. More precisely, Eq.~\eqref{eq:Hnn} becomes
\begin{equation}
    \left(R^pH\right)_{,w}
    =\frac{\epsilon p}{2}\left(R^{p-1}-\mathcal Q^2R^{-p-1}\right),
    \label{eq:RNn-boundary-H}
\end{equation}
where
\begin{equation}
    \mathcal Q^2:=\frac{\kappa Q^2}{p(p+1)\Omega_{n-2}}.
    \label{eq:RNn-calQ-def}
\end{equation}
The boundary value of $Z$ is not an independent data. It follows directly from the definition of $Z$, Eq.~\eqref{eq:Zn}, and is given by
\begin{equation}
    Z(w,0)=2\epsilon R^pR_{,w}=2\epsilon R^p r_N.
    \label{eq:RNn-Z-boundary}
\end{equation}

Before solving the non-extremal and extremal branches, it is useful to recall the geometric meaning of the horizon conditions that will be imposed below. On each null hypersurface $w=\mathrm{constant}$, the spherical sections have induced metric $\gamma_{AB}=r^2q_{AB}$, and hence $\sqrt{\gamma}=r^{p+1}\sqrt{q}$. The expansion of the congruence generated by $\ell^a=\partial_\lambda$ is therefore
\begin{equation}
    \theta_{(\ell)}=\frac{1}{\sqrt{\gamma}}\mathcal L_\ell\sqrt{\gamma}=(p+1)\frac{r_{,\lambda}}{r}.
\end{equation}
Thus $r_{,\lambda}=0$ is equivalent to $\theta_{(\ell)}=0$. In a stationary Reissner-Nordström geometry, the outer null hypersurface with vanishing outgoing expansion is identified with the future event horizon $\mathcal H^+$.

There is also a complementary interpretation associated with the hypersurfaces $\lambda=\mathrm{constant}$. Such a hypersurface becomes null whenever $V=0$. On it, the vector $\partial_w$ is null because $g_{ww}=-V$, and it is proportional to the second null vector $n^a$ defined in Eq.\eqref{eq:null-frame}.
Indeed, on $V=0$, one has $n^a\partial_a=-\epsilon\partial_w$, and the corresponding expansion reduces to
\begin{equation}
    \theta_{(n)}\Big|_{V=0}=-(p+1)\epsilon\frac{r_{,w}}{r}.
\end{equation}
Therefore, on a null hypersurface satisfying $V=0$, the condition $r_{,w}=0$ is equivalent to $\theta_{(n)}=0$. A two-sphere where both $r_{,\lambda}=0$ and $r_{,w}=0$ on $V=0$ has both null expansions vanishing. This is the natural bifurcation sphere in the non-extremal case. In contrast, in the extremal branch the future horizon is still characterized by the condition $r_{,\lambda}=0$ on a null hypersurface $w=\mathrm{constant}$, but there is no bifurcation sphere.

\subsubsection{Non-extremal branch}

We first consider the non-extremal branch. We shall focus on the case $\epsilon=-1$; the case $\epsilon=1$ follows analogously. In this case the horizon is expected to arise from a null hypersurface with non-zero surface gravity, and this is reflected in the choice of data at the initial two-sphere $\Sigma_0=\mathcal N_0\cap\mathcal B$. We take
\begin{equation}
    R(0)=r_0,
    \qquad
    R_{,w}(0)=0,
    \qquad
    H(0)=H_0\neq0,
    \qquad
    Q(0)=Q.
\end{equation}
The condition $R_{,w}(0)=0$, together with the boundary equation $R_{,ww}=0$, implies that $R$ remains constant along $\mathcal B$, namely $R(w)=r_0$. Equation~\eqref{eq:RNn-boundary-H} then reduces to a first-order equation for $H(w)$,
\begin{equation}
    H_{,w}=-\varkappa_0,
    \qquad
    \varkappa_0=\frac{p}{2r_0}\left(1-\frac{\mathcal Q^2}{r_0^{2p}}\right),
\end{equation}
and therefore
\begin{equation}
    H(w)=H_0-\varkappa_0w.
\end{equation}
The non-expanding hypersurface is located by $H(w_H)=0$, namely
\begin{equation}
    w_H=\frac{H_0}{\varkappa_0}.
\end{equation}
Since this hypersurface has vanishing expansion along the null generators, we identify it with $\mathcal H^+$. Its areal radius is
\begin{equation}
    r_+:=R(w_H).
\end{equation}
Since $R(w)=r_0$ for the present choice of boundary data, this gives
\begin{equation}
    r_0=r_+,
    \qquad
    \varkappa_0=\varkappa_+,
    \label{eq:sf}
\end{equation}
where
\begin{equation}
    \varkappa_+=\frac{p}{2r_+}\left(1-\frac{\mathcal Q^2}{r_+^{2p}}\right)\neq0.
\end{equation}
Thus the areal radius obtained from the first hypersurface equation is
\begin{equation}
    r(w,\lambda)=r_+ + H(w)\lambda,
    \qquad
    H(w)=H_0-\varkappa_+ w.
\end{equation}
The electromagnetic potential is determined next. Equation~\eqref{eq:RN-hier-alpha} gives
\begin{equation}
    \alpha(w,\lambda)=-\frac{ Q}{pH(w)}\left(\frac{1}{r^p}-\frac{1}{r_+^p}\right),
    \label{eq:alpha-nonext-general}
\end{equation}
where $\alpha(w,0)=0$ was imposed. Since $R_{,w}=0$, $Z(w,0)=0$, and Eq.~\eqref{eq:RN-hier-Z} gives
\begin{equation}
    Z(w,\lambda)=\frac{1}{H(w)}
    \left[r^p-r_+^p+\mathcal Q^2\left(r^{-p}-r_+^{-p}\right)\right].
    \label{eq:Z-nonext}
\end{equation}
This expression immediately produces the standard Reissner-Nordström metric function through the combination $HZ/r^p$:
\begin{equation}
    \frac{HZ}{r^p}=f(r),
    \label{eq:HZ-nonext}
\end{equation}
where
\begin{equation}
    f(r)=1-\frac{\mu}{r^p}+\frac{\mathcal Q^2}{r^{2p}},
    \qquad
    \mu=r_+^p+\frac{\mathcal Q^2}{r_+^p}.
    \label{eq:f-nonext}
\end{equation}
In particular,
\begin{equation}
    f(r_+)=0,
    \qquad
    f'(r_+)=2\varkappa_+.
\end{equation}
Thus $\varkappa_+$ is precisely the surface gravity associated with the outer horizon. Using Eq.~\eqref{eq:HZ-nonext}, Eq.~\eqref{eq:RN-hier-V} reduces to
\begin{equation}
    V_{,\lambda\lambda}=f''(r(w,\lambda)).
\end{equation}
This equation can be integrated by noting that for a fixed $w$
\begin{equation}
    \frac{\partial}{\partial\lambda}f'(r)=H(w)f''(r).
\end{equation}
The equation for $V$ can then be integrated once as
\begin{equation}
    V_{,\lambda}(w,\lambda)
    =\frac{1}{H(w)}\left[f'(r)-f'(r_+)\right],
\end{equation}
where the integration constant has been fixed by the boundary condition
$V_{,\lambda}(w,0)=0$ and by $r(w,0)=r_+$. Integrating once more,
\begin{equation}
    V(w,\lambda)=\frac{1}{H(w)}\int_0^\lambda\left[f'(r(w,\lambda'))-f'(r_+)\right]d\lambda' .
\end{equation}
and changing variables from $\lambda'$ to $r'=r(w,\lambda')$, with
$d r'=H(w)d\lambda'$, gives
\begin{equation}
    V(w,\lambda)=\frac{1}{H(w)^2}\int_{r_+}^{r}\left[f'(r')-f'(r_+)\right]dr' .
\end{equation}
Hence the solution satisfying $V=V_{,\lambda}=0$ at $\mathcal B$ is
\begin{equation}
    V(w,\lambda)=\frac{f(r)-f(r_+)-f'(r_+)(r-r_+)}{H(w)^2}
    =\frac{f(r)-2\varkappa_+(r-r_+)}{H(w)^2}.
    \label{eq:V-nonext-general}
\end{equation}
Since $r-r_+=H(w)\lambda$, this can also be written as
\begin{equation}
    V(w,\lambda)=\lambda^2\frac{f(r)-2\varkappa_+(r-r_+)}{(r-r_+)^2}.
\end{equation}
This form makes the regularity at the non-expanding hypersurface transparent. Indeed, the numerator has a second-order zero at $r=r_+$, and therefore the apparent singularity at $H(w_H)=0$ is removable. In particular,
\begin{equation}
    V(w_H,\lambda)=\frac{1}{2} f''(r_+)\lambda^2.
\end{equation}
The coordinates $(w,\lambda)$ provide the charged generalization of the 
coordinates originally introduced by 
Israel for the Schwarzschild metric\cite{Israel1966Schwarzschild}. 
In the Schwarzschild case, these 
coordinates have the remarkable 
property of forming a complete 
coordinate system covering the four 
regions of the maximal Kruskal 
extension; see also \cite{GKMMP} for a 
more general discussion. The 
electromagnetic potential is also 
regular at $w=w_H$ because the numerator in Eq.~\eqref{eq:alpha-nonext-general} is $O(r-r_+)=O(H\lambda)$.

It remains only to place the non-expanding hypersurface at a convenient coordinate location. We translate the origin of the null coordinate so that $w=w_H$ is mapped to $w=0$, and then relabel the translated coordinate again by $w$. With this convention,
\begin{equation}
    H(w)=-\varkappa_+w.
\end{equation}
The non-extremal affine-null fields take the final form
\begin{align}
    r(w,\lambda)&=r_+ -\varkappa_+w\lambda,
    \\
    V(w,\lambda)&=\frac{f(r)-2\varkappa_+(r-r_+)}{\varkappa_+^2w^2},
    \\
    A_a dx^a&=\frac{Q}{p\varkappa_+w}\left(\frac{1}{r^p}-\frac{1}{r_+^p}\right)dw.
\end{align}
Finally, let us relate this affine-null form to the standard Eddington-Finkelstein representation $$(u,r,x^A)$$. The required transformation follows from
\begin{equation}
    d\lambda=\frac{1}{H}\left(dr-r_{,w}dw\right),
    \qquad
    du=\frac{dw}{H(w)}.
\end{equation}
Thus, one has
\begin{equation}
    u=-\frac{1}{\varkappa_+}\ln|w|.
\end{equation}
Using Eq.~\eqref{eq:HZ-nonext}, the metric becomes
\begin{equation}
    ds^2=-f(r)du^2-2\,du\,dr+r^2q_{AB}dx^A dx^B.
\end{equation}
This is the outgoing Eddington-Finkelstein coordinate (with $u$ a Bondi time); recovering in this way the well-known expression for the $n$-dimensional version of the Reissner-Nordström metric.

\subsubsection{Extremal branch}

We now turn to the extremal branch in the outgoing formulation of the characteristic value problem $\epsilon=-1$. In this case the surface gravity of the outer horizon vanishes. From Eq.~\eqref{eq:sf}, this condition implies
\begin{equation}
    \mathcal Q^2=r_+^{2p},
    \label{eq:RNn-extremality-calQ}
\end{equation}
or equivalently
\begin{equation}
    Q^2=\frac{p(p+1)\Omega_{n-2}}{\kappa}r_+^{2p}.
\end{equation}
It is important to treat this branch separately. Although the extremal solution is obtained by imposing zero surface gravity, the affine-null form of the extremal geometry is not obtained by simply taking the limit $\varkappa_+\to0$ of the non-extremal solution. Instead, the hierarchy must be integrated with boundary data adapted to the extremal horizon.

We choose the data at $\Sigma_0$ as
\begin{equation}
    R(0)=r_0,
    \qquad
    R_{,w}(0)=r_N\neq0,
    \qquad
    H(0)=H_0,
    \qquad
    Q(0)=Q,
    \label{eq:RNn-ext-data}
\end{equation}
where $H_0=r_{,\lambda}|_{\Sigma_0}$. The boundary gauge is again
\begin{equation}
    \alpha|_{\mathcal B}=0,
    \qquad
    V|_{\mathcal B}=0,
    \qquad
    V_{,\lambda}|_{\mathcal B}=0.
\end{equation}
The condition $R_{,ww}=0$ then implies that the areal radius on the boundary evolves linearly with $w$:
\begin{equation}
    R(w)=r_0+r_Nw.
    \label{eq:RNn-R-ext-general}
\end{equation}
The remaining boundary equation determines $H(w)$. Namely, Eq.~\eqref{eq:RNn-boundary-H} gives
\begin{equation}
    \left(R^pH\right)_{,w}=-\frac{ p}{2}\left(R^{p-1}-\mathcal Q^2R^{-p-1}\right),
    \label{eq:RNn-ext-H-boundary}
\end{equation}
which integrates to
\begin{equation}
    R^pH=r_0^pH_0-\frac{1}{2r_N}
    \left[R^p-r_0^p+\mathcal Q^2\left(R^{-p}-r_0^{-p}\right)\right].
    \label{eq:RNn-ext-H-integrated}
\end{equation}
The location of the horizon is now determined geometrically. We require the existence of a null hypersurface $w=w_H$ on which the expansion of the generators vanishes, namely
\begin{equation}
    r_{,\lambda}=H(w_H)=0.
\end{equation}
If the areal radius of this hypersurface is denoted by $r_+:=R(w_H)$, then Eq.~\eqref{eq:RNn-R-ext-general} gives
\begin{equation}
    w_H=\frac{r_+-r_0}{r_N}.
    \label{eq:RNn-ext-wH}
\end{equation}
It is useful to introduce the extremal metric function
\begin{equation}
    f_+(r)=\left[1-\left(\frac{r_+}{r}\right)^p\right]^2.
    \label{eq:RNn-fplus}
\end{equation}
Evaluating Eq.~\eqref{eq:RNn-ext-H-integrated} at $w=w_H$ gives
\begin{equation}
    H_0=-\frac{1}{2r_N}\left[1-\left(\frac{r_+}{r_0}\right)^p\right]^2
    =-\frac{1}{2r_N}f_+(r_0).
    \label{eq:RNn-ext-H0-condition}
\end{equation}
This relation expresses the initial value $H_0$ in terms of the position of the non-expanding null hypersurface. With Eq.~\eqref{eq:RNn-extremality-calQ}, Eq.~\eqref{eq:RNn-ext-H-integrated} becomes
\begin{equation}
    H(w)=-\frac{1}{2r_N}\left[1-\left(\frac{r_+}{R(w)}\right)^p\right]^2
    =-\frac{1}{2r_N}f_+(R(w)).
    \label{eq:RNn-H-ext-general}
\end{equation}
The first hypersurface equation then gives the areal radius in the simple affine form
\begin{equation}
    r(w,\lambda)=R(w)+H(w)\lambda.
    \label{eq:RNn-r-ext-general}
\end{equation}
The electromagnetic potential follows next from Eq.~\eqref{eq:RN-hier-alpha}. Imposing the boundary gauge $\alpha|_{\mathcal B}=0$, one obtains
\begin{equation}
    \alpha(w,\lambda)=-\frac{ Q}{pH(w)}\left(r^{-p}-R^{-p}\right)
    =\frac{2r_NQ}{p f_+(R)}\left(\frac{1}{r^p}-\frac{1}{R^p}\right).
    \label{eq:RNn-alpha-ext-general-fplus}
\end{equation}
The boundary value of $Z$ is fixed by its definition, Eq.~\eqref{eq:Zn}, and therefore is not an independent data:
\begin{equation}
    Z(w,0)=-2r_NR^p.
    \label{eq:RNn-Z-ext-boundary}
\end{equation}
Integrating Eq.~\eqref{eq:RN-hier-Z} with $\mathcal Q^2=r_+^{2p}$ yields
\begin{equation}
    Z(w,\lambda)=2\epsilon r_NR^p+\frac{1}{H(w)}
    \left[r^p-R^p+r_+^{2p}\left(r^{-p}-R^{-p}\right)\right].
    \label{eq:RNn-Z-ext}
\end{equation}
Substituting Eq.~\eqref{eq:RNn-H-ext-general} into this expression gives the expected extremal Reissner-Nordström combination,
\begin{equation}
    \frac{HZ}{r^p}=f_+(r).
    \label{eq:RNn-HZ-ext}
\end{equation}
This identity also allows one to read off the Bondi mass. Indeed, using Eq.~\eqref{eq:MS-affine} and the relation $g^{ab}r_{,a}r_{,b}=HZ/r^p$, Eq.~\eqref{eq:RNn-HZ-ext} gives
\begin{equation}
    M=\frac{(p+1)\Omega_{n-2}}{2\kappa}r^p\left[1-f_+(r)\right]
    =\frac{(p+1)\Omega_{n-2}}{2\kappa}\left(2r_+^p-\frac{r_+^{2p}}{r^p}\right).
\end{equation}
Taking $r\to\infty$ gives
\begin{equation}
    m_B=\frac{(p+1)\Omega_{n-2}}{\kappa}r_+^p,
\end{equation}
and combining this with Eq.~\eqref{eq:RNn-extremality-calQ},
\begin{equation}
    m_B^2=\frac{(p+1)\Omega_{n-2}}{\kappa p}Q^2
    =\frac{(n-2)\Omega_{n-2}}{\kappa(n-3)}Q^2.
\end{equation}
The same identity simplifies the last hypersurface equation. Using Eq.~\eqref{eq:RNn-HZ-ext}, Eq.~\eqref{eq:RN-hier-V} reduces to
\begin{equation}
    V_{,\lambda\lambda}=f_+''(r(w,\lambda)).
\end{equation}
The solution satisfying the boundary conditions is
\begin{equation}
    V(w,\lambda)=\frac{f_+(r)-f_+(R)-f_+'(R)(r-R)}{H(w)^2}
    =\frac{4r_N^2}{f_+(R)^2}\left[f_+(r)-f_+(R)-f_+'(R)(r-R)\right].
    \label{eq:RNn-V-ext-fplus-general}
\end{equation}

For completeness, and also to make the static character of the solution explicit, we display the corresponding Killing vector. Let
\begin{equation}
    \xi=A(w,\lambda)\partial_w+B(w,\lambda)\partial_\lambda .
\end{equation}
The relevant Killing equations are
\begin{eqnarray}
    (\mathcal L_\xi g)_{\lambda\lambda}
    &=&-2\,\partial_\lambda \xi^w=0,
    \\
    (\mathcal L_\xi g)_{\lambda w}
    &=&-\partial_w\xi^w-\partial_\lambda\xi^\lambda
    -V\partial_\lambda\xi^w=0.
\end{eqnarray}
Since the static Killing field must preserve the areal radius, it must satisfy $\xi(r)=0$. Using $r=R(w)+H(w)\lambda$ and $R_{,w}=r_N$, one obtains
\begin{equation}
    B=-\frac{A}{H}\left(r_N+H_{,w}\lambda\right).
\end{equation}
The first Killing equation gives $A=A(w)$, and the second gives
\begin{equation}
    A_{,w}-A\frac{H_{,w}}{H}=0.
\end{equation}
Thus $A(w)=C_\xi H(w)$. Choosing the normalization $C_\xi=1$, the Killing vector is
\begin{equation}
    \xi=H(w)\partial_w-\left[r_N+H_{,w}(w)\lambda\right]\partial_\lambda.
    \label{eq:RNn-xi-ext-general}
\end{equation}
Using Eq.~\eqref{eq:RNn-H-ext-general}, this becomes
\begin{equation}
    \xi=-\frac{1}{2r_N}f_+(R)\partial_w-
    \left[r_N-\frac{1}{2}f_+'(R)\lambda\right]\partial_\lambda.
    \label{eq:RNn-xi-ext}
\end{equation}
Its norm is
\begin{equation}
    g(\xi,\xi)=-H^2V+2 H\left(r_N+H_{,w}\lambda\right)=-\frac{HZ}{r^p}=-f_+(r),
\end{equation}
as expected for the standard static Reissner-Nordström Killing vector.

As in the non-extremal branch, it is convenient to translate the origin of the null coordinate so that the non-expanding hypersurface $w=w_H$ is placed at $w=0$. After relabelling the translated coordinate by $w$, the regular extremal affine-null fields are
\begin{align}
    r(w,\lambda)&=R(w)-\frac{1}{2r_N}f_+(R(w))\lambda,
    \qquad
    R(w)=r_+ + r_Nw,
    \label{eq:RNn-final-ext-r}
    \\
    V(w,\lambda)&=\frac{4r_N^2}{f_+(R(w))^2}
    \left[f_+(r)-f_+(R(w))-f_+'(R(w))(r-R(w))\right],
    \label{eq:RNn-final-ext-V}
    \\
    A_a dx^a&=\frac{2r_NQ}{p f_+(R(w))}
    \left(\frac{1}{r^p}-\frac{1}{R(w)^p}\right)dw.
    \label{eq:RNn-final-ext-A}
\end{align}
Finally, as in the previous case, the relation with the usual extremal Reissner-Nordström-Tangherlini line element in Eddington-Finkelstein coordinates $(u,r_{\rm EF},x^A)$ is obtained through
\begin{equation}
    r_{\rm EF}=r(w,\lambda),
    \qquad
    du=\frac{dw}{H(w)}.
\end{equation}
Equivalently, in the shifted coordinate,
\begin{equation}
    R=r_+ + r_Nw,
    \qquad
    H(w)=-\frac{1}{2r_N}f_+(R),
\end{equation}
so that
\begin{equation}
    du=-2\frac{dR}{f_+(R)},
    \qquad
    \lambda=\frac{r_{\rm EF}-R}{H(w)}=-\frac{2r_N}{ f_+(R)}\left(r_{\rm EF}-R\right).
\end{equation}
With these relations, the metric takes the standard form
\begin{equation}
    ds^2=-f_+(r_{\rm EF})du^2-2\,du\,dr_{\rm EF}+r_{\rm EF}^2q_{AB}dx^A dx^B,
\end{equation}
and the potential is gauge-equivalent to
\begin{equation}
    A_a dx^a\simeq -\frac{ Q}{p r_{\rm EF}^p}du.
\end{equation}
For $n=4$, namely $p=1$,
\begin{equation}
    f_+(R)=\left(\frac{r_Nw}{r_+ + r_Nw}\right)^2,
\end{equation}
and
\begin{equation}
    r(w,\lambda)=r_+ + r_Nw-\frac{ r_Nw^2}{2(r_+ + r_Nw)^2}\lambda,
\end{equation}
which reproduces the regular four-dimensional extremal Israel-type form. This expression was also obtained in \cite{GKMMP}, both by directly solving the null geodesic equations and by using an alternative hierarchical affine-null system. Before ending this section, let us note that, without loss of generality, one may set $r_N=-1$; see the discussion in \cite{Madler:2025ibn}.

\section{Summary}

In this paper, we have discussed the affine-null hierarchical formulation of the Einstein-Maxwell system coupled to a charged complex scalar field in $n$-dimensional spherical symmetry. The introduction of the auxiliary variables $Q$, $Z$ and $L$ restores a sequential hypersurface hierarchy, allowing the fields $r$, $Q$, $\alpha$, $Z$, $L$ and $V$ to be obtained by radial integration once the scalar data and boundary data are specified. We derived the corresponding asymptotic initial value problem, including the scalar news, the Bondi mass and the asymptotic charge balance laws. We also analyzed the local vertex formulation and showed how regularity fixes the leading behavior of all fields near the center. Finally, in the scalar-free sector, we recovered the non-extremal and extremal Reissner-Nordström-Tangherlini black hole solutions directly from the hierarchy.

We would like to emphasize that this hierarchical system can be easily generalized to massive scalar fields, in that case, the { equations of the CIBVP for $F\in\{\Phi$, $r_{,\lambda}$, $Z$, $V\}$} 
acquire additional terms such as {$F\rightarrow F+A^{(F)}_k r^k m^2|\Phi|^2$ with $k\ge 0$ while the equation for $\mathcal{L}$ acquires a term proportional to $m^2\Phi.$}

As a non-trivial check of the formulation, we studied the scalar-free sector and recovered the $n$-dimensional Reissner-Nordström-Tangherlini family directly from the hierarchical equations. Both the non-extremal and extremal branches were obtained in affine-null form. In the non-extremal case, the horizon is located by the vanishing of $r_{,\lambda}$ on a constant-$w$ null hypersurface and the resulting solution is regular after shifting the origin of $w$ to the future horizon. In the extremal case, the hierarchy must be integrated separately, rather than obtained as a naive zero surface gravity limit of the non-extremal coordinates. The final affine-null fields were shown to reduce to the standard Eddington-Finkelstein form by the corresponding coordinate transformation. 

One important open problem is whether the present $n$-dimensional charged system admits a conformal affine-null formulation with variables that remain regular at future null infinity. In four dimensions, for the massless scalar field with Maxwell coupling, such a formulation can be constructed by introducing conformal variables and auxiliary fields that lead to a hierarchy regular at $\mathscr I^+$ \cite{madler-gallo-conf}. Moreover, that conformal system is closely related to the system obtained by compactifying the physical affine parameter and regularizing the corresponding physical variables. In the present case the situation appears less direct. The conformal weights of the scalar field depend on the spacetime dimension, the Maxwell sector is not conformally invariant in the same way outside four dimensions, and the charge terms introduce additional inverse powers of the areal radius. Thus, a conformal completion of the $n$-dimensional Einstein-Maxwell-scalar affine-null hierarchy would require a separate analysis and, most likely, a different choice of regular variables at $\mathscr I^+$. 

A further application of the system developed here is the study of critical collapse in higher dimensions. The affine-null formulation is particularly well suited for this purpose because the hypersurface equations remain regular at surfaces where the outgoing expansion vanishes, allowing one to track the evolution across apparent horizons and into the black-hole interior. This feature has already proved useful in four-dimensional simulations of supercritical scalar-field collapse in affine-null coordinates, where the apparent horizon can be followed up to the approach to the final singularity\cite{MadlerBaakeHosseiniWinicour2024}. The present $n$-dimensional charged system provides a natural framework to revisit Choptuik-type scaling, echoing and universality in higher dimensions, now including electric charge and a complex scalar field. It would also allow comparison with previous studies of higher-dimensional scalar collapse, self-similar solutions and critical exponents, as well as with recent investigations of complex and massive scalar-field collapse. 

These directions suggest that the affine-null hierarchy derived here is not only a useful analytic representation of known black-hole solutions, but also a practical starting point for numerical studies of dynamical charged collapse. In particular, it provides a framework for studying the formation and evolution of charged black holes, the approach to extremality, the behavior of apparent horizons, and the possible dependence of critical phenomena on the spacetime dimension and on the gauge charge of the scalar field.

\section*{Acknowledgements}

 E.G. and T.M gratefully acknowledge the hospitality of the Institute for Computational and Experimental Research in Mathematics (ICERM) at Brown University, where part of this work was carried out. E.G. also acknowledges financial support from SeCyT-UNC and CONICET. L.B. and T.M. acknowledge support from the FONDECYT REGULAR with No. 1262002 and No. 1262349 of the Chilean national funding agency ANID.


\end{document}